\documentclass[11pt,a4paper]{article}
\usepackage[margin=2.2cm]{geometry}
\usepackage[T1]{fontenc}
\usepackage[utf8]{inputenc}
\usepackage{lmodern}
\usepackage{microtype}
\usepackage{graphicx}
\usepackage{booktabs}
\usepackage{array}
\usepackage{tabularx}
\usepackage{longtable}
\usepackage{ragged2e}
\usepackage{needspace}
\usepackage{amsmath}
\usepackage{pdflscape}
\usepackage[backend=biber,style=authoryear,natbib=true,maxcitenames=2]{biblatex}
\usepackage[hidelinks]{hyperref}
\usepackage{xurl}
\usepackage{caption}
\usepackage{setspace}
\title{Reproducibility, Methodological Robustness, and Synthetic-Domain Selection Properties of AFINO for TESS Stellar-Flare QPP Analysis}
\author{Marc Balboa Corominas\\\small Independent Researcher, Spain\\\small \href{mailto:m.balboacorominas.research@gmail.com}{m.balboacorominas.research@gmail.com} (ORCID: 0009-0006-7571-9193)}
\date{}
\begin{document}
\maketitle
\begin{abstract}
Quasi-periodic pulsation (QPP) searches in stellar flares depend on methodological choices that can affect both classification and the interpretation of recovered periods. A defensible assessment therefore requires explicit separation of reproducibility, robustness, and known-truth performance. We characterize the reproducibility, methodological robustness, numerical behavior, and synthetic-ground-truth selection properties of a frozen public AFINO implementation in the TESS context while keeping observational reference evidence separate from known synthetic truth. The frozen baseline reproduced five published TESS detections in the initial reproduction cohort. A ten-event observational pilot and a catalogue-scale robustness experiment then tested prospectively specified temporal-window and processing perturbations. At catalogue scale, 51 of 61 published-QPP references did not reproduce their frozen reference state under the F3A baseline, so robustness claims were conditioned on baseline-concordant references rather than interpreted as accuracy. In a separately developed synthetic programme, the untouched AFINO 0.5 rule was evaluated once on an independent HELDOUT set: 152 of 1800 synthetic positives were selected and no false selections were observed among 1800 synthetic nulls, with finite Wilson uncertainty. The resulting 156-stratum selection surface and period-recovery summaries remain conditional on the frozen synthetic domain and, for period recovery, on selection; they therefore provide a bounded characterization of methodological robustness and synthetic-domain selection behavior without establishing observational validation or a population-level correction.
\end{abstract}

\section{Introduction}
QPPs are recurring intensity modulations reported in solar and stellar flares, but the inference problem is methodologically difficult because flare evolution, colored noise, finite observing windows, detrending, gaps, and model choice can all produce or suppress apparent periodic structure. The literature represented in the frozen Bibliographic Audit II (BAII) includes Fourier- and wavelet-based significance testing, automated model comparison, blind methodological benchmarks, and machine-learning classifiers. That diversity motivates treating a QPP catalogue entry not as a self-authenticating physical label but as an output of a specified analysis procedure whose reproducibility and robustness can be tested. Methodological diversity and significance-testing concerns are represented in the frozen corpus by \citet{Inglis2016}, \citet{Pugh2017}, and \citet{Broomhall2019}.

AFINO is an automated model-comparison framework that evaluates competing power-spectral models and uses Bayesian information criterion differences to identify a localized excess consistent with a QPP component. Large-scale AFINO applications predate the present programme, and the frozen BAII corpus also contains catalogue-scale TESS QPP work and neural-network classification. Accordingly, the contribution evaluated here is not framed as a first-ever catalogue, catalogue-scale AFINO study, or injection-recovery experiment. Its focus is the combined prospective architecture linking public-code reproduction, observational robustness, numerical diagnostics, development-only rule assessment, and independent synthetic held-out characterization. Catalogue-scale and machine-learning precedents include \citet{HowardMacGregor2022}, \citet{Belov2024}, \citet{Joshi2025}, and \citet{Wang2025}.

The observational starting point is the published 20-second-cadence TESS flare analysis that motivated the reproduction cohort. Related TESS studies demonstrate that QPP catalogue construction and classification are already active methodological areas, while adjacent flare and QPP work contains injection-recovery and known-truth evaluation analogues. These precedents make two boundaries important. First, observational reference labels cannot be promoted to physical ground truth merely because they were published. Second, synthetic detection performance cannot be transported to the observed TESS population without explicit assumptions and independent evidence. The broader solar--stellar context is also summarized in the frozen recent review by \citet{Reale2026}.

BAII itself is an auxiliary positioning plane rather than an experimental result. Its search was prospectively bounded, source access was uneven, and literature after the frozen search window is not systematically covered. We therefore use the audit only to constrain wording and comparator claims. In particular, the manuscript makes no global priority claim and does not treat non-retrieval of a precedent as evidence that no precedent exists. External comparators identified by the audit were considered in the design record but were not executed head-to-head in the frozen scientific programme.

Within those boundaries, the manuscript asks a narrower methodological question: what can be established when one fixed public AFINO implementation is followed from observational reproduction through controlled robustness tests to synthetic known-truth evaluation, without collapsing the different truth conditions? The resulting evidence is deliberately layered. The study can describe whether published reference classifications reproduce, whether classifications change under prospectively specified perturbations, whether binary decisions are stable across external optimizer seeds, and how the frozen rule behaves under designed synthetic truth. It cannot, from these evidence classes alone, establish observational physical truth or a population correction.

A central practical difficulty is that apparently similar QPP detections can arise from pipelines with different temporal support, preprocessing, stochastic-background assumptions, and decision rules. A catalogue-level agreement or disagreement is therefore ambiguous unless the underlying implementation and input contract are specified. The frozen literature audit supports this methodological framing without implying that one existing technique is preferred. It also motivates the prospective strategy used here: define the analysis choices before examining the corresponding robustness or held-out outcome, then preserve failures and inadmissible inputs rather than editing the design around the observed result.

\clearpage
\section{Evidence and analysis architecture}
The analysis is organized into five primary evidence planes with distinct truth conditions. F0 is observational reproduction: it asks whether a frozen public-code protocol can reproduce selected published TESS reference classifications. F1 is a synthetic/numerical benchmark with known truth inside a designed domain. F2 is a small observational robustness pilot in which repeated methodological perturbations are applied to frozen reference events. F3A extends that robustness design to catalogue scale but retains observational reference labels rather than physical truth. F3B returns to known synthetic truth, separates DEVELOPMENT from an independent single-use HELDOUT split, and evaluates a rule frozen before HELDOUT access. BAII sits beside these planes as documentary positioning evidence rather than as a sixth scientific result.

\begin{figure}[htbp]
\centering
\includegraphics[width=0.98\linewidth]{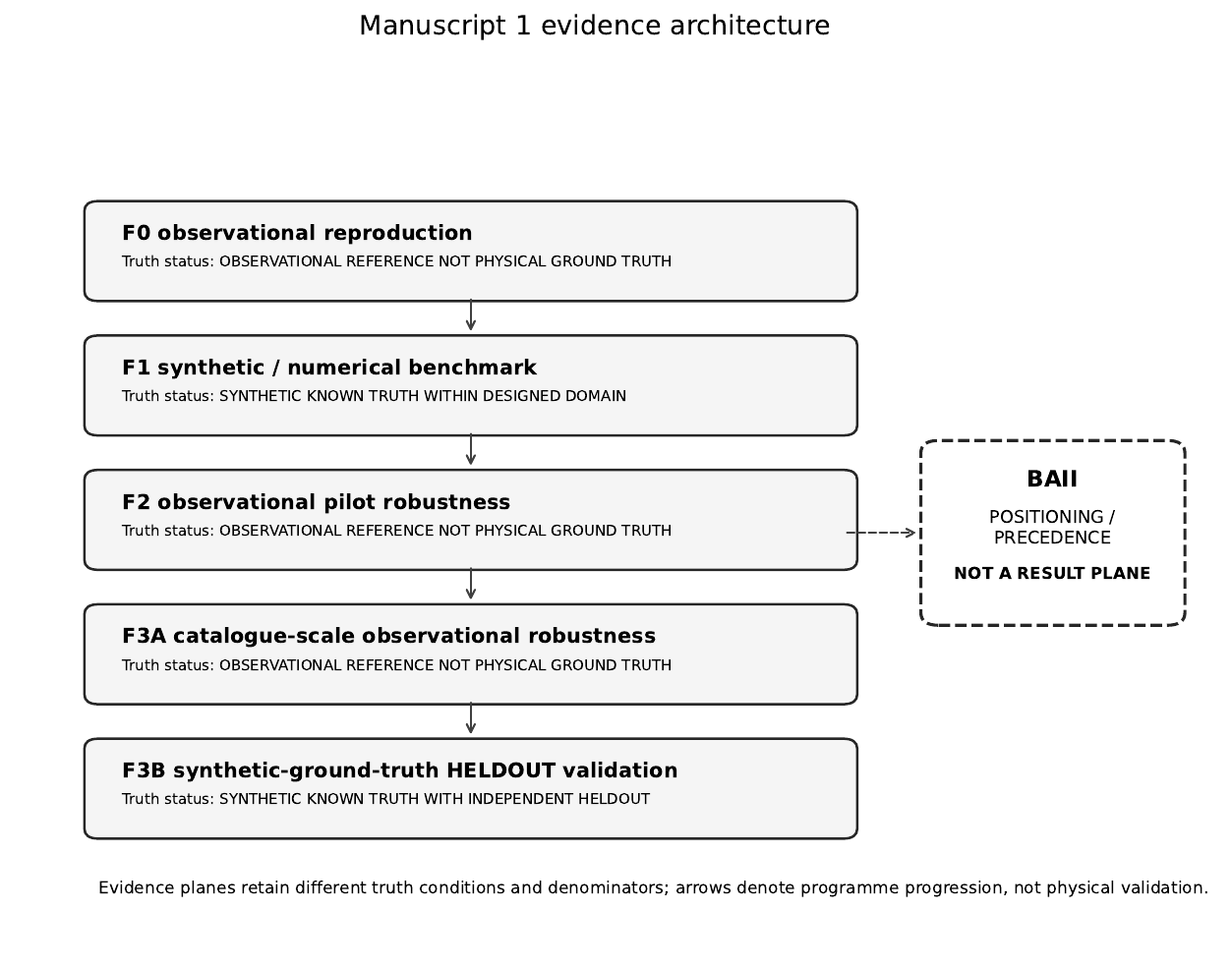}
\caption{Evidence architecture of the programme. The five primary scientific planes retain their own truth conditions; BAII is an auxiliary literature-positioning evidence rather than a result plane.}
\label{fig:m1f01}
\end{figure}

This separation governs the terminology used throughout the paper. Reproduction means agreement with a frozen observational reference under a specified public-code protocol; it does not mean physical validation. Robustness means stability or change of a reference classification under repeated perturbations; it does not imply a confusion matrix when truth is unknown. Synthetic-ground-truth performance refers only to the designed simulation domain. Finally, held-out validation is used only with an explicit synthetic-ground-truth qualifier. The word validation is not used as a generic umbrella for all phases.

\Needspace{8\baselineskip}
\begin{table}[htbp]
\centering
\caption{Evidence planes and their interpretation boundaries.}
\label{tab:evidence_planes}
\small
\begin{tabularx}{\textwidth}{@{}>{\RaggedRight\arraybackslash}p{1.15cm}>{\RaggedRight\arraybackslash}p{3.25cm}>{\RaggedRight\arraybackslash}X>{\RaggedRight\arraybackslash}X@{}}
\toprule
Plane & Data / truth status & What it establishes & What it does not establish \\
\midrule
F0 & Observational reference; not physical ground truth. & Reproducibility of the frozen public AFINO baseline on the observational reproduction cohort and the operational baseline used downstream. & Observational physical QPP truth, population sensitivity/specificity/FPR, or identity with unpublished private analysis code. \\
F1 & Synthetic known truth; designed domain. & Controlled synthetic-domain benchmark behavior, support dependence, period behavior and numerical diagnostics of the frozen AFINO implementation. & Observational population performance or transport to real TESS prevalence. \\
F2 & Observational reference; not physical ground truth. & Pilot-scale sensitivity of observational reference classifications to prospectively frozen methodological perturbations. & Classifier accuracy, observational FPR/sensitivity/specificity, or physical truth. \\
F3A & Observational reference; not physical ground truth. & Catalogue-scale baseline reproduction limitations, methodological classification robustness, seed stability and conditional period robustness. & Observational validation, sensitivity, specificity, FPR, physical truth, or a selection function. \\
F3B & Synthetic known truth; independent HELDOUT. & Independent single-use HELDOUT synthetic-ground-truth performance, final synthetic selection function and conditional period recovery for the frozen baseline. & Observational validation, real-TESS prevalence, physical truth or a validated population correction. \\
BAII (aux.) & Literature-positioning evidence; non-result plane. & Bounded literature-positioning, overlap and precedence constraints under the frozen BAII search/extraction scope. & A scientific result of the AFINO/TESS experiment or unrestricted global priority. \\
\bottomrule
\end{tabularx}
\end{table}

The evidence-plane architecture also determines which denominators can be compared. F2 and F3A contain repeated methodological variants nested within observational events; their transition counts cannot be pooled with synthetic realization counts. DEVELOPMENT and HELDOUT share a simulation design but have different evidential roles and are reported separately. Period-recovery samples are conditional subsets of classification samples. By carrying these distinctions into the manuscript structure, the analysis avoids a common failure mode in which convenient numerical summaries obscure that the underlying observations answer different scientific questions.

\clearpage
\section{Methods}
\subsection{Frozen AFINO baseline and observational reproduction}
The observational baseline used the public AFINO release at the frozen repository commit and package version 0.5. The effective TESS protocol was reconstructed empirically before downstream robustness analyses. Primary inputs used PDCSAP flux with finite values retained, an inclusive cadence-index window anchored to the nearest time markers, and time expressed relative to the first included cadence. The public preprocessing normalized by the mean, applied a Hann window, computed the Fourier spectrum, and compared three spectral models: a power law plus constant, the same background plus a localized Gaussian component, and a broken power law plus constant.

The QPP decision required both BIC contrasts to exceed the frozen threshold: the QPP model had to improve on the simple background and on the broken-power-law alternative. The operational rule was strict, with both delta-BIC quantities greater than 10. Candidate periods were constrained to the empirically reconstructed 40--300 s range. External optimizer seeds were reset before each model fit and used to characterize decision stability, while formal optimizer convergence remained unauditable because the public analysis path did not retain the optimizer success/message state.

The F0 reproduction cohort was intentionally small and confirmatory. A calibration case was used to reconstruct the effective windowing convention, after which four additional positive/reference pairs were selected without inspecting their light curves or AFINO outcomes. The final F0 evidence therefore comprised five published QPP detections and five paired events that had not been selected as QPPs in the source catalogue. Reproduction was defined as recovery of the frozen reference decision under the specified public baseline; the exercise did not attempt to reconstruct the authors' unavailable private TESS adapter or establish physical truth.

Model competition is important to the interpretation of the baseline. Selection is not triggered merely because the localized QPP component fits better than a simple colored-noise background; it must also outperform the broken-power-law alternative. The second comparison is intended to reduce the risk that broadband spectral curvature is mislabeled as a localized oscillatory excess. Because the public implementation and the reconstructed TESS input protocol are both frozen, downstream changes in classification can be attributed to the predefined methodological perturbations within the project rather than to an evolving baseline implementation.

\subsection{Synthetic benchmark}
F1 introduced a controlled synthetic benchmark to separate known-truth behavior from observational reproduction. The primary grid contained stationary QPP-present and null flare realizations with 20 s sampling, red-noise slopes, several time-series lengths, QPP periods, and relative QPP amplitudes fixed prospectively by the benchmark design. The primary condition table contained 111 conditions, including 12 null conditions and 99 QPP-present conditions. Each primary condition used 40 realizations for the selection summary. These quantities define the designed domain; they are not estimates of the distribution of real TESS flares.

A second, nested-support experiment changed the observed prefix length within parent realizations. Its purpose was to examine the total consequence of extending temporal support, not to isolate a causal ``number of bins'' effect. Extending a prefix changes the realized noise segment, the normalization, Hann weighting, and Fourier grid simultaneously. The nested analysis therefore treated support extension as a compound methodological perturbation. Primary decision rates remained fully stratified, and selected-period summaries were kept separate from the formal center of the QPP model when that model did not satisfy the selection rule.

Numerical diagnostics were recorded independently of the binary classifier. The frozen analyses tracked model bounds, warnings, variation of BIC across external optimizer seeds, and multiplicity flags for the broken-power-law model. A stable binary decision across seeds was not treated as evidence of a unique global optimum. Likewise, a warning or bound contact was not used post hoc to accept or reject a scientific decision. This separation was maintained because the public implementation did not expose a complete convergence record that would support a formal global-convergence claim.

The synthetic benchmark was used primarily as a map of operating behavior rather than as a single score. Conditions were examined in their frozen strata so that support, signal period, signal fraction, and red-noise properties remained visible. This mattered because the selection rule requires simultaneous evidence against two backgrounds and the limiting comparison can change with the observed support. The nested experiment further showed why a longer segment cannot be interpreted as a clean intervention on sample size: the additional samples alter several ingredients of the spectral estimate at once. The benchmark therefore motivates robustness tests without claiming causal identification of individual preprocessing effects.

\subsection{Pilot observational robustness}
F2 translated the synthetic motivation into a prospective observational robustness pilot. The frozen cohort contained 10 events organized as five pairs, with one published-QPP reference and one matched not-selected reference per pair. Each event was evaluated over 13 temporal windows and six processing profiles, yielding the frozen perturbation matrix per event. The profiles covered finite-flux, quality-filtered, SAP/PDCSAP, detrending, and quality-zero variants defined before execution. Repeated variants were nested within events and were never treated as independent physical labels.

Input admissibility was an explicit outcome. Variants that violated sampling, cadence-count, or peak-retention requirements were labeled \texttt{INPUT\_INADMISSIBLE} rather than recoded as non-selections. Classification transitions were defined relative to a frozen W00/P00 baseline for each event, while additional temporal and processing contrasts used their own local references. The complete W00--W12-style temporal perturbation set and P00--P05 processing profiles are machine-readable in \path{foundation/f0-f2/phase2/fase2_tarea01_window_perturbations.csv} and \path{foundation/f0-f2/phase2/fase2_tarea01_processing_profiles.csv}, with the external-seed grid and eligibility rules frozen in \path{fase2_tarea01_observational_robustness_preregistration.json}. Period robustness was computed only when both compared states remained selected and had an available selected-period output. Seed stability used the frozen external seed grid on eligible baseline-window variants.

The F2 analysis treated input admissibility as part of methodological robustness because preprocessing can change whether the AFINO input contract is satisfied. A quality filter can remove a flare peak, a shortened window can leave too few cadences, and filtering can create irregular sampling. Recording these cases explicitly prevents a methodological perturbation from appearing artificially conservative simply because problematic inputs disappear from the analyzed denominator. The same principle later became important at catalogue scale and in the design of the synthetic-ground-truth evaluation plane.

\subsection{Catalogue-scale robustness}
F3A scaled the same prospective 13 x 6 robustness architecture to 122 observational references, split evenly between published-QPP references and published not-selected references. The full plan therefore contained 9516 methodological variants. Before transition analysis, every event passed through a baseline-reproduction gate using W00/P00 and external seed 0. Only baseline-concordant references were eligible for retained/lost or retained/gained transition interpretation. Baseline mismatch and input inadmissibility were preserved as separate states rather than being forced into the transition matrix.

This baseline gate was central to the semantics of F3A. For QPP references, a transition could be called selected-retained or selection-lost only if the reference itself reproduced under the frozen baseline. For not-selected references, a transition could be called not-selected-retained or selection-gained under the same baseline-concordance requirement. These categories are repeated methodological transitions of observational reference states. They are deliberately not named true positives, false negatives, true negatives, or false positives.

The catalogue-scale numerical-stability plane was again separated from methodological transitions. Input-eligible W00/P00 events were rerun across a frozen external-seed grid, and the binary decision was summarized by event. Distinct fitted parameter payloads, warnings, bound contacts, and unavailable formal convergence diagnostics remained numerical evidence rather than classification truth. Period robustness was separately restricted to baseline-comparable selected-to-selected variants; no period was imputed to lost, not-selected, or inadmissible states.

F3A was therefore analyzed in two stages: reference reproduction first, perturbation robustness second. This ordering was not cosmetic. If the frozen baseline does not reproduce an event's reference state, then a subsequent perturbed classification cannot be described coherently as retained or lost relative to that reference without changing the meaning of the transition. The explicit gate ensures that transition labels remain conditional on a reproduced starting point. Mismatch and inadmissibility remain scientifically visible outcomes in their own right, rather than being absorbed into the robustness categories.

\subsection{Synthetic injection--recovery and synthetic-ground-truth held-out validation}
F3B was designed to answer the known-truth performance question that F2 and F3A could not answer. DEVELOPMENT and HELDOUT were separated prospectively, with the final decision rule frozen before HELDOUT truth was consumed. The classifier plane used synthetic QPP-present and synthetic QPP-absent series that were eligible for AFINO under the frozen design. The untouched baseline rule remained delta-BIC01 greater than 10 and delta-BIC21 greater than 10 with strict comparison. All performance statements in this phase are therefore synthetic-ground-truth statements within the preregistered domain. The primary generator combined a fixed asymmetric flare envelope with a Gaussian Fourier-synthesis power-law background, $P(f)\propto f^{-\alpha}$, standardized to a sample standard deviation of 0.005. Positive series added a stationary envelope-modulated sinusoid, $qE(t)\sin[2\pi(t-t_{\rm peak})/P+\phi]$, with $q\in\{0.01,0.02,0.04\}$, phase uniform on $[0,2\pi)$, and period sampled log-uniformly from 40--300 s subject to at least three cycles; null series omitted this component. No independent photometric white-noise term was added in the primary generator.

Threshold development was confined to DEVELOPMENT. A single preregistered candidate emerging from that search improved balanced accuracy and sensitivity relative to the baseline but incurred a large specificity loss. Promotion required all four frozen criteria to pass. The candidate passed three criteria but failed the specificity-preservation criterion, so it was not promoted. Runner-up rescue, alternate post hoc candidate searches, and threshold retuning after HELDOUT were prohibited. Consequently, HELDOUT evaluated the original 10/10 baseline rather than a rule chosen from HELDOUT behavior.

The independent HELDOUT split was single-use. Once the frozen baseline metrics, stratified selection surface, and conditional period-recovery outputs were produced, the split was considered consumed and unavailable for later threshold development. Classification metrics used known synthetic truth; the selection surface retained structural no-exposure cells instead of treating them as zero selection; and period recovery was computed only for selected true positives with finite recovered periods. These safeguards prevent HELDOUT from serving simultaneously as a synthetic-ground-truth evaluation resource and a new development resource.

The separation between DEVELOPMENT and HELDOUT was designed so that a failed preregistered promotion decision remained informative rather than being bypassed. DEVELOPMENT could support threshold exploration under the preregistered policy, but a candidate could reach HELDOUT only if every promotion criterion was satisfied. Once the candidate failed those criteria, the protocol prohibited searching for a more favorable substitute. This preserved the original baseline as the only admissible held-out target. The later HELDOUT result therefore evaluates a rule whose identity is independent of the held-out outcomes, which is essential for the interpretation of the synthetic-ground-truth evaluation plane.

\section{Results}
\subsection{Baseline reproduction and synthetic benchmark behavior}
The F0 public-code baseline reproduced all five published QPP detections in the frozen reproduction cohort and preserved the non-selected status of the five paired comparison events. In the positive reference events, the double-BIC decision was stable across the external seed checks used in the reproduction phase. This establishes an operational public baseline that can be reproduced on the examined cohort. It does not establish that the public implementation is documentarily identical to the authors' unavailable private TESS pipeline, nor that the observational labels are physical ground truth.

The F1 synthetic benchmark showed strong domain dependence. In the primary benchmark, M1 was not selected in any of 480 null realizations, whereas the 99 QPP-present conditions ranged from no selections to complete selection within individual strata. Seventy-eight positive conditions had zero selections and 21 had at least one. All positive conditions at the two shortest tested supports, N=15 and N=30, remained at zero selections; selection became concentrated in longer series, larger QPP fractions, and shorter injected periods. These are properties of the frozen synthetic generator and AFINO protocol, not observational performance estimates.

The nested-support experiment reinforced the conclusion that temporal support cannot be reduced to a monotonic ``more data is better'' rule. Across positive trajectories, both upward and downward threshold crossings occurred. Most positive trajectories were never selected, two became selected at an intermediate support and then reverted, and two appeared only at the longest tested support. The preregistered contrast intended to distinguish improvement relative to the simple and broken-power-law backgrounds was negative in most adjacent positive transitions, so the dominant pattern did not support a simple monotonic support hypothesis.

The F1 outcomes also show why a marginal synthetic detection fraction would be misleading. Detection was concentrated in particular combinations of support and injected signal properties, while other regions of the designed domain produced no selections. In addition, support extension was non-monotonic in the nested trajectories. A single aggregate rate would hide both structures and could suggest a smooth performance characteristic that the frozen experiments do not establish. The manuscript therefore uses the synthetic benchmark as evidence of domain dependence and numerical behavior, leaving the final stratified selection characterization to the independently held-out F3B plane.

\subsection{Pilot robustness}
The F2 pilot contained 780 planned variants across the 10 frozen events. Of these, 514 were eligible for AFINO and 266 were input-inadmissible. The inadmissible states were retained in the planned denominator rather than recoded as non-selections. Among the eligible variants, the event-level baseline comparison produced 140 selected-retained variants, 136 selection losses, 238 not-selected-retained variants, and no baseline-relative selection gains. The presence of both retained and lost selections in the published-QPP reference role shows dependence of the reference classification on methodological choice within the pilot.

Temporal-window and processing contrasts also contained local classification changes, confirming that the global baseline summary did not conceal all direction changes. Because each event contributed repeated windows and processing profiles, these transition counts are not independent trials. The pilot therefore supports a qualitative robustness conclusion: prospectively specified methodological choices can change the reference classification of individual observational events. It does not provide observational sensitivity, specificity, or false-positive rates, and the paired not-selected role is not a physically verified negative class.

\begin{figure}[htbp]
\centering
\includegraphics[width=0.98\linewidth]{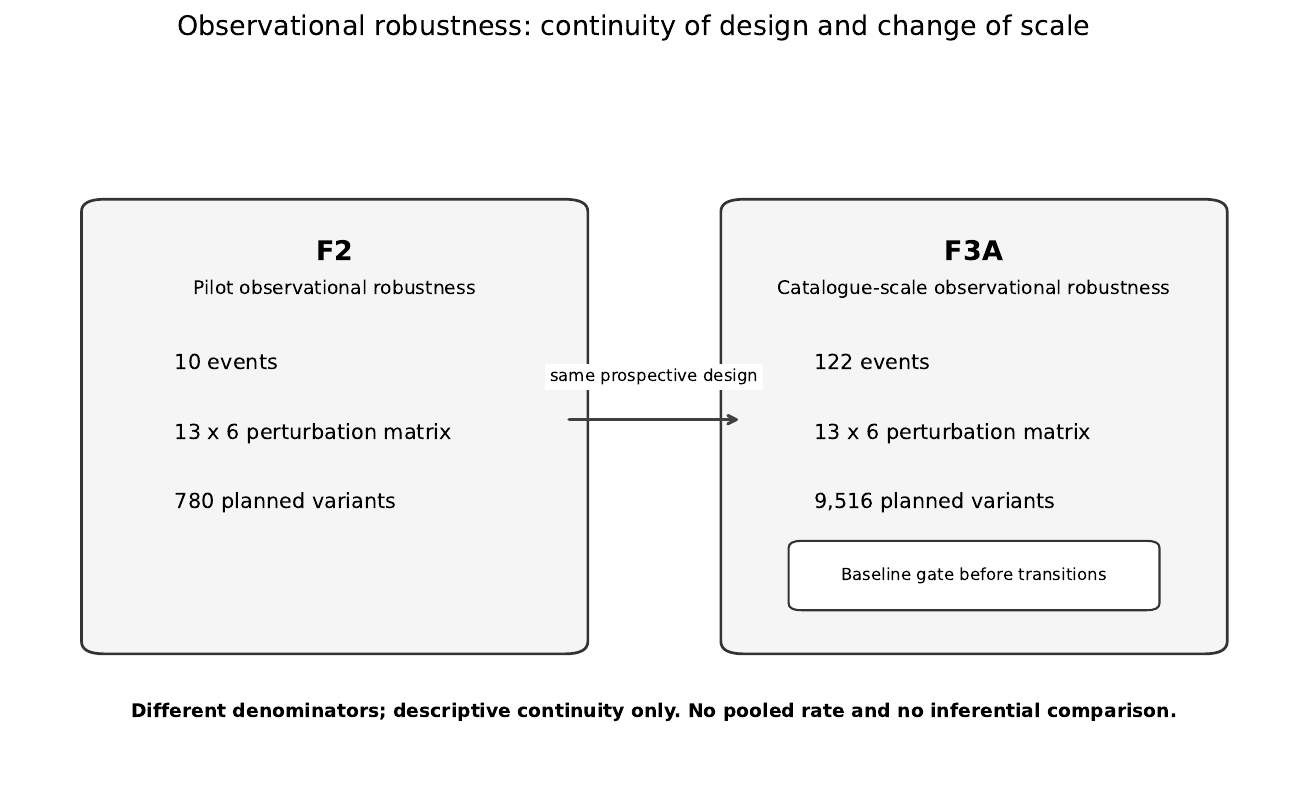}
\caption{Progression from the F2 observational pilot to the catalogue-scale F3A stress test. Denominators remain phase-specific and are not pooled.}
\label{fig:m1f02}
\end{figure}

The pilot's local temporal and processing comparisons reinforce the baseline-relative transition summary. Some variants changed classification relative to a nearby window or processing reference even when they did not create a gain relative to the global W00/P00 baseline. This distinction matters because a gain is meaningful only relative to its declared reference state. The pilot therefore demonstrates event-level dependence on analysis choices while also showing why transition counts should not be detached from the comparison definition that generated them.

\subsection{Catalogue-scale robustness and baseline reproduction}
F3A expanded the robustness design to 122 observational references and 9516 planned variants, of which 6422 were eligible and 3094 were inadmissible. The baseline gate materially changed the interpretation relative to F2: only 65 references were concordant with their frozen reference state, 51 were baseline mismatches, and six were baseline-inadmissible. All mismatches occurred in the published-QPP reference role. Among 61 QPP references, eight were concordant, 51 mismatched, and two were inadmissible; among 61 not-selected references, 57 were concordant, none mismatched, and four were inadmissible.

The 51/61 mismatch result is a reproduction limitation, not a physical falsification result. It states that the frozen W00/P00/seed0 F3A baseline did not reproduce the published-QPP reference state for those events under the project input and implementation contract. F3A did not establish a cause for the mismatch, and the cause remains unresolved within that phase. Consequently, all perturbation claims were conditioned on the narrower baseline-concordant scope rather than treating the full published-QPP reference set as a known positive class.

Within the transition-eligible baseline-concordant scope, F3A contained 295 selected-retained transitions and 171 selection losses for QPP references. The not-selected reference scope contained 3178 not-selected-retained transitions and zero selection gains. These are cell-level repeated measures from the frozen perturbation matrix. The coexistence of retained and lost QPP selections supports catalogue-scale dependence of reference classifications on methodological perturbations. Conversely, the absence of selection gains is a robustness observation within a reference cohort, not an observational false-positive rate.

\Needspace{8\baselineskip}
\begin{longtable}{@{}p{0.31\textwidth}p{0.16\textwidth}p{0.45\textwidth}@{}}
\caption{F3A catalogue-scale robustness counts. Observational reference roles are not physical ground truth.}\label{tab:f3a_counts}\\\\
\toprule
Category and scope & Count/state & Interpretation \\\\
\midrule
\endfirsthead
\toprule
Category and scope & Count/state & Interpretation \\\\
\midrule
\endhead
\textbf{Cohort}\newline all F3A observational references & 122 events & Observational reference cohort; not physical ground truth \\
\textbf{Primary variants}\newline frozen 13 x 6 matrix across 122 events & 9516 & Repeated-measure methodological variants \\
\textbf{Input eligibility}\newline eligible variants & 6422 & Eligible for classification within the frozen stress test \\
\textbf{Input eligibility}\newline inadmissible variants & 3094 & Structural/methodological inadmissibility retained as outcome \\
\textbf{Baseline gate}\newline all references: concordant & 65 & 8 QPP-reference + 57 not-selected-reference \\
\textbf{Baseline gate}\newline all references: mismatch & 51 & Published-QPP reproduction mismatch; not physical falsity \\
\textbf{Baseline gate}\newline all references: inadmissible & 6 & No baseline comparison available \\
\textbf{Baseline gate}\newline QPP reference: concordant / mismatch / inadmissible & 8 / 51 / 2 & Reference-state gate only \\
\textbf{Baseline gate}\newline not-selected reference: concordant / mismatch / inadmissible & 57 / 0 / 4 & Reference-state gate only; mismatch count is not FPR \\
\textbf{Transitions}\newline QPP selected retained / selection lost & 295 / 171 & Cell-level repeated measures conditional on baseline concordance \\
\textbf{Transitions}\newline not-selected retained / selection gained & 3178 / 0 & Cell-level repeated measures; zero gain is not observational FPR \\
\textbf{Numerical stability}\newline optimizer summaries & 116 & Input-eligible W00/P00 event scope \\
\textbf{Numerical stability}\newline seed decisions & 1160 & Frozen seed grid; not evidence of unique optimizer convergence \\
\textbf{Numerical stability}\newline seed-discordant events & 0 & Binary output seed-stable in frozen scope \\
\textbf{Period robustness}\newline period-comparable variants & 295 & Conditional on retained selection and baseline comparability \\
\bottomrule
\end{longtable}

Catalogue scale changes the evidential emphasis. The perturbation matrix remains useful, but the dominant methodological finding becomes the baseline gate itself: most published-QPP references are not transition-eligible because their reference state is not reproduced. The robustness results therefore describe a narrower conditional population than the original QPP-reference cohort. By contrast, the not-selected reference role is largely baseline-concordant and remains stable under the frozen perturbations. The asymmetry is a reproducibility feature of the observational reference construction; without independent physical labels it cannot be interpreted as asymmetric classifier accuracy.

\clearpage
\subsection{Numerical stability}
F1 separated binary-decision stability from numerical uniqueness. None of the 111 primary benchmark conditions changed the binary decision across the external seeds used for the stability analysis, yet 97 of 111 conditions met the preregistered broken-power-law multiplicity criterion based on BIC variation. In the nested benchmark, the same qualitative separation persisted: binary decisions were seed-stable in the examined stability conditions while M2 multiplicity remained common. Frequent bounds and M2 warnings were therefore retained as numerical diagnostics rather than converted into post hoc exclusion rules.

F3A provided a larger observational stability check. All 116 input-eligible W00/P00 events preserved their binary classification across the frozen seed grid, corresponding to 1160 seed decisions and zero seed-discordant events. At the same time, each event retained multiple fitted parameter payloads and formal convergence remained unauditable. The supported conclusion is therefore seed stability of the binary output in this frozen scope, not convergence to a unique numerical solution.

Conditional period robustness in F3A was evaluated only where the baseline and perturbed variant were both selected and comparable. The frozen table contained 295 such rows. This quantity is not a catalogue-wide period-accuracy denominator: selection losses, not-selected states, and input-inadmissible variants do not contribute a recovered-period comparison. The period plane therefore complements, rather than replaces, the classification-robustness plane.

Across F1 and F3A, the numerical story is consequently one of decision stability coexisting with parameter multiplicity. This is useful because it narrows what optimizer variability appears to affect in the tested scopes: the external seed does not alter the binary decision, even though fitted parameter payloads and diagnostics vary. The converse claim is not warranted. Seed stability does not prove that the objective surface has a single optimum, that every run reached a global solution, or that individual fitted parameters have a unique physical interpretation. Those stronger questions remain outside the auditable numerical evidence.

\Needspace{10\baselineskip}
\begin{figure}[htbp]
\centering
\includegraphics[width=0.98\linewidth]{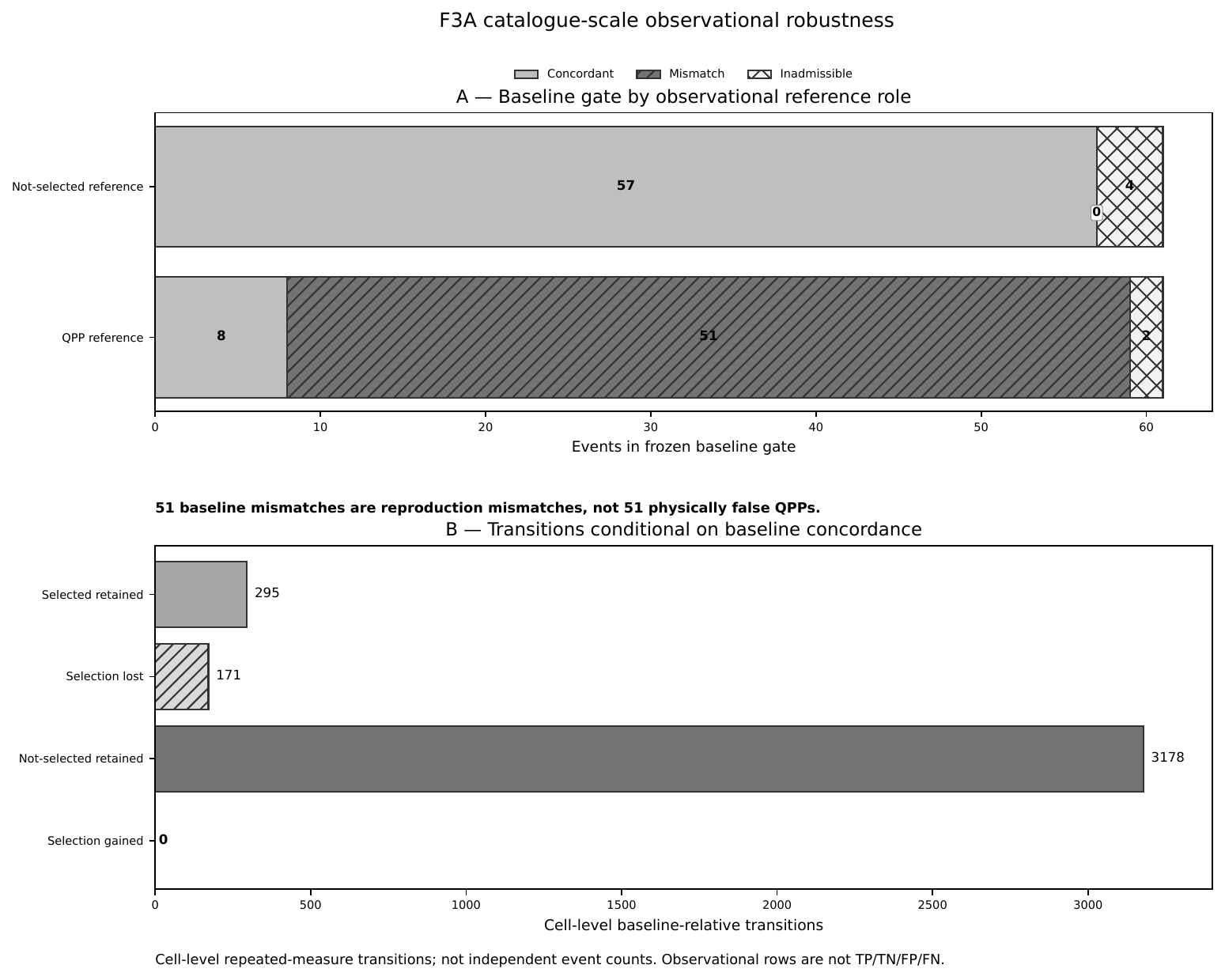}
\caption{Catalogue-scale F3A baseline gate and baseline-conditioned methodological transitions. The mismatch category is a reproduction mismatch, and transition labels are not confusion-matrix outcomes.}
\label{fig:m1f03}
\end{figure}

\clearpage
\subsection{DEVELOPMENT synthetic performance and rule gate}
Under the frozen 10/10 baseline, DEVELOPMENT contained 143 true positives, 1657 false negatives, 1799 true negatives, and one false positive among 3600 primary classifier rows. The corresponding synthetic-domain sensitivity was 0.07944, specificity 0.99944, and observed false-positive fraction 0.000556. These are split-specific known-truth metrics. They are not observational sensitivity, specificity, or FPR, and they are not pooled with HELDOUT.

The DEVELOPMENT candidate rule was not promoted. Although its balanced accuracy exceeded the baseline and three of four promotion criteria passed, the lower confidence limit for the candidate-minus-baseline specificity difference failed the preregistered preservation requirement. The preregistered promotion rule therefore retained the untouched baseline as the final HELDOUT target: delta-BIC01 greater than 10 and delta-BIC21 greater than 10. No runner-up search or alternative rescue was performed.

The candidate-rule episode illustrates the value of a multidimensional promotion criterion. A rule that improves sensitivity and balanced accuracy can still be scientifically unattractive if it sacrifices too much specificity for the intended operating profile. Because specificity preservation was a prospectively required criterion rather than an editorial preference applied after the fact, its failure had a deterministic consequence: the candidate was not promoted. The result is therefore not a comparison between an optimized rule and the held-out baseline; it is evidence that the predefined development path did not establish a replacement rule suitable for independent synthetic-ground-truth validation.

\subsection{Independent HELDOUT performance}
On the independent synthetic HELDOUT, the frozen baseline produced 152 true positives, 1648 false negatives, 1800 true negatives, and zero false positives. With 1800 synthetic positives and 1800 synthetic nulls, the HELDOUT sensitivity was 0.08444 and the specificity point estimate was 1.0. The observed false-positive fraction was 0/1800. The standard Wilson interval for that zero count retained a non-zero upper bound of approximately 0.00213, so the observed absence of false selections does not establish a population FPR of exactly zero.

DEVELOPMENT and HELDOUT thus show the same broad operating profile when described separately: low synthetic sensitivity and extremely high synthetic specificity for the frozen baseline. The split values are close descriptively, but the programme did not pool them or run an equivalence test, and no inference of observational transport follows. In particular, the HELDOUT zero false-selection count is a finite-sample observation in a designed null distribution; it is not evidence that false selection is impossible in either the broader synthetic population or TESS observations.

\Needspace{8\baselineskip}
\begin{table}[htbp]
\centering
\caption{Synthetic-ground-truth performance of the frozen baseline. Confidence intervals are Wilson intervals where defined.}
\label{tab:synthetic_performance}
\small
\begin{tabular}{@{}lrrrr@{}}
\toprule
Metric & DEVELOPMENT & 95\% Wilson & HELDOUT & 95\% Wilson \\\\
\midrule
TP & 143 &  & 152 &  \\
FN & 1657 &  & 1648 &  \\
TN & 1799 &  & 1800 &  \\
FP & 1 &  & 0 &  \\
Sensitivity & 0.079444 & [0.067828, 0.092852] & 0.084444 & [0.072467, 0.098191] \\
Specificity & 0.999444 & [0.99686, 0.999902] & 1 & [0.99787, 1] \\
FPR & 0.000556 & [9.808e-05, 0.00314] & 0 & [2.168e-19, 0.00213] \\
Balanced accuracy & 0.539444 &  & 0.542222 &  \\
\bottomrule
\end{tabular}
\par\vspace{0.4em}\footnotesize All quantities are split-specific synthetic-ground-truth metrics within the frozen design and are not observational performance estimates.
\end{table}

The HELDOUT confusion matrix should be read jointly rather than through the specificity point estimate alone. The positive class shows that the baseline selects only a small fraction of the injected QPPs in the designed domain, while the null class shows that false selections were not observed in this finite sample. Wilson intervals preserve uncertainty on both quantities. This combination is consistent with a conservative operating point under the synthetic design, but the result does not determine whether the same trade-off holds under observational noise, flare morphologies, gaps, or prevalence.

\subsection{Synthetic selection function and conditional period recovery}
The final HELDOUT empirical selection surface contains 156 strata. It preserves the frozen stratification over sample support, red-noise slope, QPP fraction, and period-bin structure, with null strata represented separately. Nine combinations are marked \texttt{STRUCTURAL\_NO\_EXPOSURE} rather than assigned a zero selection estimate. No DEVELOPMENT rows are pooled into this surface, and no smoothing, interpolation, regression, or probabilistic selection model is fitted. The product is therefore a direct stratified summary of the independent frozen HELDOUT domain.

Among the 152 selected true positives, recovered periods generally remained close to the injected values. This describes period localization conditional on successful selection, not overall detection completeness: the same HELDOUT contains 1800 eligible synthetic positives, of which only 152 were selected.

The stratified surface and period-recovery product add structure that is absent from the headline confusion matrix. The selection surface shows which regions of the frozen simulation design contribute observed exposure and how selection varies within those exposed cells. The period table then asks a different conditional question for successfully selected positives. Keeping structural no-exposure distinct from an observed zero and keeping period recovery distinct from classifier coverage prevents two different missing-data mechanisms from being collapsed into apparently poor performance.

\clearpage
\begin{landscape}
\begin{figure}[p]
\centering
\includegraphics[width=0.99\linewidth,height=0.84\textheight,keepaspectratio]{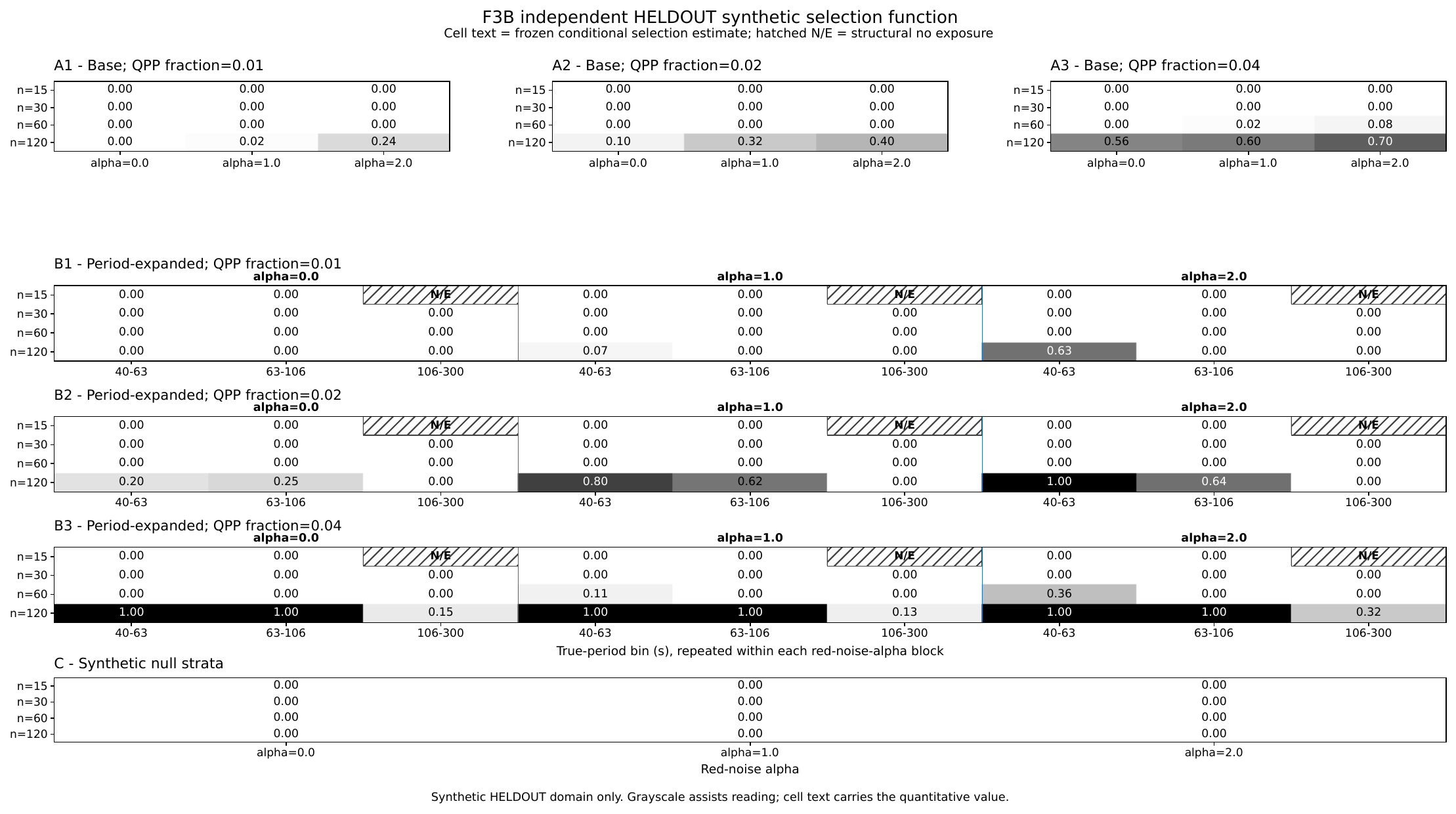}
\caption{Independent HELDOUT synthetic selection surface across the frozen design. Hatched N/E cells denote structural no exposure rather than zero selection.}
\label{fig:m1f04}
\end{figure}
\end{landscape}
\clearpage

\begin{figure}[htbp]
\centering
\includegraphics[width=0.98\linewidth]{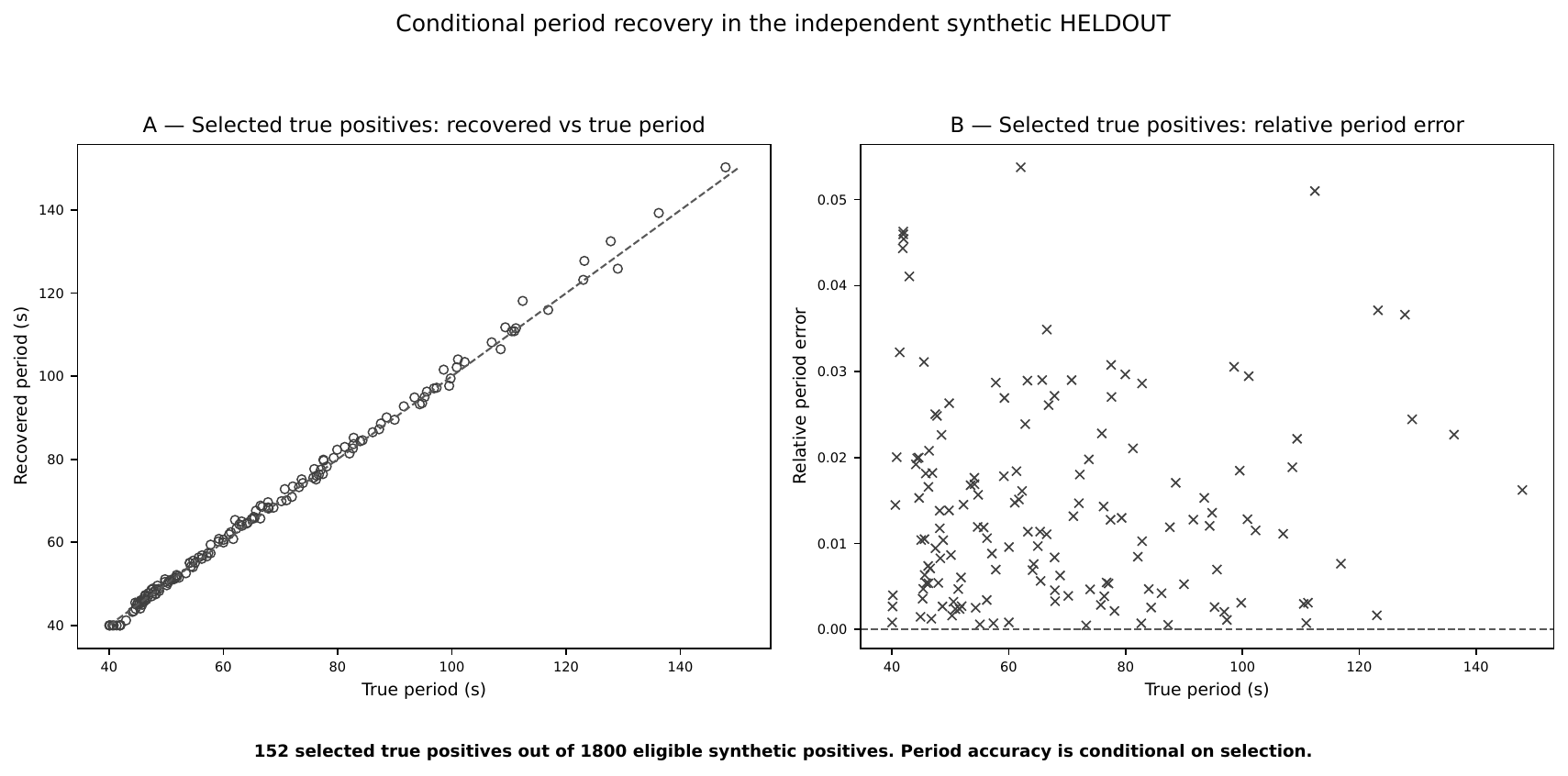}
\caption{Conditional period recovery for the 152 selected true positives out of 1800 eligible synthetic positives. Period accuracy is conditioned on selection and is not a completeness measure.}
\label{fig:m1f05}
\end{figure}

\section{Discussion}
\subsection{Methodological robustness}
The strongest cross-phase robustness result is qualitative continuity rather than a pooled rate. The F2 pilot and F3A catalogue-scale experiment used the same prospectively specified window/profile logic, preserved inadmissibility as an explicit outcome, and both showed that QPP-reference classifications can be retained or lost under methodological perturbation. At catalogue scale, dependence of reference classifications on the frozen analysis choices remained evident. At the same time, F3A exposed a baseline-reproduction limitation that was not present in the small F2 cohort, demonstrating why robustness cannot be assessed responsibly without first checking whether the reference baseline itself reproduces.

The F0 5/5 and F3A 8/61 reproduction results are not estimates of the same quantity. F0 was a small confirmatory reproduction exercise that included the calibration case used to reconstruct effective windowing, whereas F3A applied the frozen baseline prospectively to a distinct catalogue-scale reference cohort.

Two forms of stability should also be distinguished. Binary decisions were stable across the frozen external-seed checks in the examined F3A scope, but fitted parameter solutions were not unique and convergence was not formally auditable. Separately, F3B shows that a rule can be numerically and procedurally frozen yet still have low sensitivity under known synthetic truth. Reproducibility of execution, seed stability of a decision, robustness to methodological perturbation, and classification performance are therefore different properties. Treating them as synonyms would overstate what any individual phase establishes.

Several safeguards restrict the reported scope. Input-inadmissible cases remain explicit; F3A transitions are interpreted only after baseline reproduction; seed stability is separated from optimizer convergence; and DEVELOPMENT is isolated from HELDOUT evaluation. These restrictions reduce some denominators but make the assumptions behind each result explicit.

\subsection{Selection-limited behavior}
The held-out operating profile is selection-limited in the frozen synthetic domain. The important result is not simply that no HELDOUT null was selected, but that this very-high-specificity behavior coexists with low positive recovery. The finite Wilson interval around the zero false-selection count is a reminder that a point estimate at a boundary is not certainty about an underlying rate. The programme therefore avoids the shorthand ``perfect specificity'' and instead reports the observed count, denominator, uncertainty, and domain.

Selection conditioning also changes how period performance should be read. Period localization among selected true positives can be good even when most injected positives are not selected at all. This is not a contradiction: classification and parameter recovery answer different conditional questions. A method may identify the period accurately after a strong event crosses the model-comparison thresholds while failing to select many weaker or less favorable events. For population inference, the missing selected fraction would matter at least as much as the conditional period errors.

The low-sensitivity/high-specificity synthetic profile also has an important practical implication for interpreting selected samples. A catalogue produced by a conservative threshold may contain events whose fitted QPP periods are well localized conditional on selection, yet still represent only a small and strongly conditioned subset of the underlying positive signals in the simulation domain. Scientific conclusions based only on selected events can therefore differ from conclusions about detection completeness. The present study does not correct for that conditioning; it documents it so that later population work must confront the selection process explicitly.

\subsection{Observational versus synthetic evidence}
The F3A mismatch result illustrates why observational-reference and synthetic-ground-truth planes must remain distinct. Fifty-one published-QPP references fail the frozen F3A baseline reproduction gate, but the data do not reveal which component of the upstream catalogue/pipeline difference is responsible. Calling those events ``false QPPs'' would replace a reproducibility statement with a physical-truth claim that the experiment cannot support. The appropriate response is to narrow subsequent robustness claims to reproduced baselines and retain the mismatch itself as an important methodological limitation.

The same logic applies to the absence of F3A selection gains. The not-selected references are observational roles, and their repeated perturbations are not an independently verified negative population. Zero gains therefore do not define an observational FPR. Conversely, F3B has known synthetic truth and can support synthetic sensitivity and false-selection metrics, but those metrics do not automatically describe real TESS data. The programme consequently provides reproduction evidence, observational robustness evidence, and synthetic-ground-truth evidence without claiming observational validation of AFINO.

Observational evidence can test reproducibility without knowing physical truth, whereas synthetic evidence can test classification against known truth without establishing observational realism. The observational phases reveal implementation and preprocessing dependencies that a simplified generator may miss, while the synthetic phase supplies confusion-matrix semantics that observational reference labels cannot provide. Their evidential value is complementary rather than interchangeable.

\subsection{Why no correction is claimed}
A population correction is not claimed for two independent reasons. First, the DEVELOPMENT threshold candidate failed its preregistered specificity-preservation gate, so no new classifier rule was promoted to synthetic-ground-truth held-out evaluation; HELDOUT evaluated the untouched baseline. Second, the final selection surface is conditional on a designed synthetic population rather than an observational target population. Applying that surface directly to TESS occurrence rates would require assumptions about how the simulation dimensions map onto real flare morphology, noise, data quality, cadence structure, and the distribution of unmodelled effects. Those transport assumptions were not established here.

Because the preregistered promotion criteria were not met, the present analysis does not support an observational population correction. The HELDOUT false-selection result and conditional period recovery therefore remain synthetic-domain evidence and are not used as a post-hoc correction.

\Needspace{8\baselineskip}
\begin{longtable}{@{}>{\RaggedRight\arraybackslash}p{0.20\textwidth}>{\RaggedRight\arraybackslash}p{0.38\textwidth}>{\RaggedRight\arraybackslash}p{0.34\textwidth}@{}}
\caption{Reader-facing interpretation boundaries.}\label{tab:claim_boundaries}\\
\toprule
Finding & Supported interpretation & Key limitation \\
\midrule
\endfirsthead
\toprule
Finding & Supported interpretation & Key limitation \\
\midrule
\endhead
F0 reproduction & The frozen public AFINO baseline reproduced the five published TESS detections in the F0 cohort. & Reproduction is not observational validation or physical confirmation; paper-code identity remains unresolved. \\
F3A 51/61 baseline mismatch & Under W00/P00/seed0, 51 of 61 published-QPP references did not reproduce their frozen reference state. & These are reproduction mismatches, not physical falsifications; the cause is unresolved within F3A and reference labels are not truth. \\
F3A zero selection gains & No baseline-relative selection gains were observed in the eligible baseline-concordant not-selected reference scope. & Repeated observational references are not known negatives; this is not an observational FPR or specificity estimate. \\
F3A seed stability (116/116) & All 116 input-eligible W00/P00 events preserved binary classification across the frozen seed grid. & Decision stability does not establish unique optimizer convergence. \\
F3B HELDOUT FP=0 (0/1800) & No false selections were observed among 1800 HELDOUT synthetic nulls, with finite-sample Wilson uncertainty. & This synthetic finite-sample result does not establish population FPR=0 or observational specificity. \\
F3B selection function (156 strata) & The 156-row HELDOUT stratified empirical selection surface characterizes the frozen synthetic domain. & No DEVELOPMENT pooling, smoothing, or observational transport was performed; it is not a universal TESS correction. \\
Conditional period recovery (152 selected TPs) & Period recovery is summarized for 152 selected true positives and can remain accurate despite low selection coverage. & The result is conditioned on selection and does not imply high completeness or classifier sensitivity. \\
DEVELOPMENT candidate-rule failure & The candidate failed the preregistered specificity-preservation criterion and was not promoted; HELDOUT evaluated the frozen 10/10 baseline. & No runner-up rescue or post-hoc threshold search was performed; HELDOUT was not used to tune thresholds. \\
Observational population correction & The Phase 3B results do not establish an observational population correction. & This limitation is part of the scientific interpretation and must remain explicit. \\
Programme evidence-plane boundary & The programme provides distinct reproduction, robustness, and synthetic-ground-truth evidence. & The evidence planes remain separate and do not establish observational validation of AFINO. \\
BAII priority constraint & The contribution is positioned around the combined prospective robustness and held-out architecture. & BAII is bounded to its frozen scope and does not support unrestricted global novelty or first-ever claims. \\
\bottomrule
\end{longtable}

A reproducible synthetic selection surface is not, by itself, a correction function for a survey. Such a correction would require a target estimand and a defensible mapping from simulated covariates and latent states to the observational population; neither was established here. The frozen F3B surface therefore does not establish an observational population correction.

\subsection{Implications for population analyses}
Future population inference requires a new transport layer rather than reuse of the consumed HELDOUT split. At minimum, such work would need to define the target observational population, justify how synthetic signal and noise dimensions approximate that target, incorporate relevant quality/gap/cadence regimes, and quantify uncertainty introduced by the transport itself. If thresholds are tuned further, they require new independent evidence because the current HELDOUT has been consumed. These requirements place observational correction and population inference beyond the evidence freeze of the present manuscript.

The same future-work boundary applies to comparator claims. BAII identified several relevant methodological families, but the frozen F3A/F3B programmes did not execute a head-to-head comparison against wavelet, machine-learning, hidden-Markov, Gaussian-process, or other candidate approaches. The present study therefore cannot rank AFINO against those methods. A later comparison could be valuable if it is prospectively scoped and evaluated under shared inputs and truth conditions, but it would be a new scientific analysis rather than an editorial extension of this manuscript.

A subsequent population programme can use the present work as a design constraint rather than as a ready-made correction. The observed failure modes identify dimensions that a transport study should consider, including baseline reproducibility, input admissibility, data-quality structure, temporal support, and the separation between selection and recovered parameters. The consumed HELDOUT cannot serve as a new optimization target, so any revised classifier or transport model requires an independently frozen evaluation resource. This preserves the evidential value of the current held-out result while allowing the programme to progress beyond it.

\section{Conclusions}
The programme establishes a reproducible public AFINO baseline for the initial TESS reference cohort and shows, first in a small pilot and then at catalogue scale, that observational reference classifications can depend on prospectively specified window and processing choices. Scaling also reveals that baseline reproduction itself is a material gate: the catalogue-scale published-QPP reference set contains many baseline mismatches, which must not be interpreted as physical falsifications.

Under known synthetic truth, the untouched 10/10 AFINO 0.5 rule displays a low-sensitivity, very-high-specificity operating profile in the frozen F3B domain. A DEVELOPMENT candidate was not promoted because it failed the preregistered specificity-preservation gate, and the independent HELDOUT remained single-use. The resulting selection surface and period-recovery products characterize that synthetic domain but do not observationally validate AFINO. The resulting synthetic characterization therefore does not establish an observational population correction.

Taken together, the evidence supports a bounded methodological conclusion: reproducibility, robustness, numerical stability, and synthetic-ground-truth performance can be characterized rigorously only when their truth conditions and denominators remain explicit. The manuscript therefore provides a documented basis for later population work while intentionally stopping short of observational validation or a population correction.

\section*{Acknowledgments}

Generative AI tools were used selectively for editorial support and technical assistance, including the review and refinement of portions of the code and the preparation of some visualizations. The research workflow, scientific analysis, validation of results, and interpretation were carried out and verified by the author, who takes full responsibility for the content of this work.

Code checks included comparisons with independently implemented numerical reference cases, deterministic replays of model outputs, checkpoint restart and idempotence tests, and recalculation of classification decisions from frozen results. Figure and table values were traced to frozen source artifacts and checked for numerical and scientific identity during manuscript preparation. Validation records and source bindings are retained in the public repository. These checks concern implementation and provenance; they do not establish observational physical validation.

\section*{Funding}
This research received no external funding.
\section*{Competing interests}
The author declares no competing interests.
\section*{Data, code, and reproducibility resources}
\begin{sloppypar}
The scientific results in this manuscript resolve to frozen repository evidence rather than to newly computed quantities. The public AFINO baseline is package version 0.5 at repository commit \texttt{6aceac9518fc8056052807e666da9d0c8bebb010}. The TESS-QPP research repository is available at \url{https://github.com/marc-balboa-corominas/tess-qpp}; manuscript archival artifacts are mirrored in the project OSF record at \url{https://osf.io/vhktr/}. The M1.4 scientific manuscript freeze is commit \texttt{10e4ac7017950f60e74a1f0fddb41f6004f7755d} (tag \texttt{manuscript1-reviewed-draft-v1}), the definitive M1.2 visual freeze is commit \texttt{7e65987511487ea7de01d1b2880cc70687823541} (tag \texttt{manuscript1-visuals-v1}), and the author-approved manuscript freeze is M1.6 commit \texttt{d1007edcdbcf98b809ed46b0810fd62148f7b2af} (tag \texttt{manuscript1-author-approved-v1}). F3A and F3B are frozen under \texttt{phase3a-complete-v2} and \texttt{phase3b-complete-v1}, respectively. Machine-readable provenance for these freezes is retained in the linked repository and OSF record. No DOI is asserted for resources for which a DOI has not been frozen.
\end{sloppypar}

\clearpage
\printbibliography
\end{document}